\documentclass[amsmath,amssymb,amsbsy,reprint,prb,preprintnumbers,showpacs,superscriptaddress]{revtex4-2}
\usepackage{graphicx,color}
\usepackage{dcolumn}
\usepackage{bm}
\usepackage{braket}
\usepackage{mathtools}
\usepackage{ulem}
\usepackage[breaklinks,colorlinks=true,linkcolor=blue,urlcolor=blue,citecolor=blue]{hyperref}
\usepackage{times}
\usepackage{physics}
\usepackage{latexsym}
\usepackage{amsmath, amssymb}
\usepackage{mathtools}
\usepackage{multirow}

\newcommand{\be}{\begin{eqnarray}}
\newcommand{\ee}{\end{eqnarray}}

\renewcommand{\theequation}{\arabic{equation}}

\def\changedd#1{\textcolor{blue}{#1}}

\begin{document}
\title{R\'{e}nyi Markov length in one-dimensional non-trivial mixed state phases and mixed state phase transitions}

\date{\today}
\author{Yoshihito Kuno} 
\affiliation{Graduate School of Engineering science, Akita University, Akita 010-8502, Japan}

\author{Takahiro Orito}
\affiliation{Department of Physics, College of Humanities and Sciences, Nihon University, Sakurajosui, Setagaya, Tokyo 156-8550, Japan}
\affiliation{Institute for Solid State Physics, The University of Tokyo, Kashiwa, Chiba, 277-8581, Japan}

\author{Ikuo Ichinose} 
\thanks{A professor emeritus}
\affiliation{Department of Applied Physics, Nagoya Institute of Technology, Nagoya, 466-8555, Japan}


\begin{abstract} 
Discovering and classifying non-trivial mixed states and mixed state phase transitions are some of the most important current issues in condensed matter and quantum information. 
In this study, we investigate some non-trivial mixed states and phase transitions between them by using the second R\'{e}nyi conditional mutual information (CMI). 
The CMI can measure mixed state ``gap'', estimated by the exponential decay rate of the second R\'{e}nyi CMI under a tripartition of system, 
which provides the second R\'{e}nyi version of the Markov length. 
We introduce an efficient numerical scheme for the calculation of the second R\'{e}nyi CMI based on the doubled Hilbert space formalism, and 
study the classification of non-trivial mixed states and the emergence of mixed state phase transitions for (i) the cluster model under odd-site local $Z$ decoherence 
and (ii) transverse field Ising model under both $ZZ$ and $X$ decoherence. 
The second R\'{e}nyi CMI is a powerful measure to study non-trivial mixed ``gapped" quantum matters and mixed phase transitions.  
In addition to this, the present study shows that the second R\'{e}nyi CMI exhibits specific behavior for the transition to strong-to-weak spontaneous symmetry breaking mixed phase.
\end{abstract}


\maketitle
\section{Introduction}
For quantum systems, noise and decoherence caused by environment are inevitable~\cite{gardiner2000}. 
Especially, for quantum memory such as toric code~\cite{Kitaev2003,dennis2002,wang2003,ohno2004} and intermediate scale quantum computer \cite{ebadi2021,bluvstein2024,Bharti2022,preskill2018}, investigation of effects of noise and/or decoherence and error correction against them have attracted significant attention in condensed matter and quantum information communities.

However, noise and decoherence do not act solely to disrupt the target physical state. 
In fact, recent developments have raised the expectations that non-trivial quantum many-body mixed states, without pure-state counterparts, 
emerge due to interactions with environment inducing decoherence, etc. 
For example, the realization pattern of symmetries in mixed states becomes rich due to effects by environment, 
such as spontaneous symmetry breaking (SSB) of strong and weak symmetries \cite{Buca_2012,groot2022}.
More precisely, theoretical study indicates the existence of distinctive non-trivial mixed state:  symmetry protected topological (SPT) phases such as average SPT (ASPT) \cite{ma2023,ma2024,Ma2024_double,Lee2025},  unconventional spontaneous symmetry breaking in mixed states induced by decoherences such as strong-to-weak spontaneous symmetry breaking (SWSSB) states~\cite{lee2023,liu2024_SSSB,guo2024,shah2024,weinstein2024,sala2024,Orito2025,kuno_2025,Lessa_PRXQ,Ando2024}, and topological order induced by decoherences such as intrinsic mixed-state topological orders~\cite{Wang_PRXQ,Fan_PRXQ,sang2024,Chen2024_v2,Sohal_PRXQuantum,Zhang2024strong,kuno_2025_v1}. The construction of concrete theoretical models and systematic exploration of such non-trivial mixed-state phases are further required, as well as their classification through the pursuit of their universal properties.

In order to classify non-trivial mixed phases, and to identify phase transitions between them, it is useful to introduce the notion of `gap' for mixed states, 
which is based on quantum informational theory. 
As a useful candidate, Markov length defined by an exponential decay rate of the conditional mutual information (CMI) was proposed in \cite{Sang2025}, 
the inverse of which can be regarded as a ``gap'' of mixed states. 
This proposal was successfully applied to SWSSB phases \cite{Lessa_PRXQ}. 

Furthermore, the Markov length can play an important role in defining mixed state phase equivalence. 
In the framework of pure ground states corresponding to a family of Hamiltonian, the state equivalence is rooted in the presence of a gap: 
If there is an adiabatic continuous parameter path between two pure ground states ``without gap closing'', these states belong to the same equivalent class.
An analogous approach for mixed states has been proposed in \cite{Sang2025}. 
If there is a continuous quantum channel map between two mixed states without closing the ``gap'', the two states can be regarded as ``states in an equivalent mixed phase''. 
This is rooted in the existence of a map called Petz map~\cite{Petz1988,Junge2018}, i.e.,
the two mixed states are invertible under the Petz map. 
The estimation of CMI and its Markov length not only introduce the notion of ``gap''(corresponding to the inverse of Markov length) of mixed states 
but also gives the phase equivalence of mixed states through the presence of a (invertible) recovery map. 
The recoverbility of a mixed state is formally and qualitatively bounded by the behavior of the CMI as shown in \cite{Sang2025}. 
If the CMIs for a mixed state on a continuous channel path between two mixed phases exhibit exponential decay, a recovery Petz map is to be constructed and 
the two mixed states can be in an equivalent mixed phase. 
Inversely, if we observe some divergence or sudden increase of the Markov length, mixed states can exhibit some singular behaviors, implying mixed state phase transitions \cite{Negari2024,Sang2025}.
Thus, the observation of the CMI for mixed states gives efficient insight into the classification of non-trivial mixed states.

Based on the usefulness of the CMI and the Markov length, we shall study an extension of the Markov length derived from the second R\'{e}nyi entropy
and its associated CMI \cite{Zhang2024strong} by applying them to two concrete systems. 
We expect that the second R\'{e}nyi CMI and its Markov length show similar behaviors to the ones of the original von Neumann CMI and its Markov length. 
The reason for focusing on the CMI defined via the second R\'{e}nyi entropy is that not only is it generally related to the conventional CMI, 
but it is also well-suited with the doubled Hilbert space formalism, enabling efficient numerical computation using matrix product states (MPS). 
The proposal of an efficient calculation way for the second R\'{e}nyi entropy through the doubled Hilbert space formalism is one of the main results of this work. 

We numerically investigate two concrete systems under decoherence and show the efficiency of the second R\'{e}nyi CMI to observe non-trivial mixed phases 
and mixed state phase transitions. 
The first target system is a cluster model under local $Z$ decoherence and the second is a transverse field Ising model (TFIM) under $ZZ$ and $X$ decoherence. 
The former system exhibits the ASPT phase for the finite decoherence, recently proposed in \cite{ma2024,Lee2025,Ma2024_double}. 
The later exhibits a SWSSB phase transition, recently founded in \cite{Orito2025}.  
For these two concrete systems, by employing the efficient numerical methods for the second R\'{e}nyi CMI, 
we demonstrate that the second R\'{e}nyi CMI and its Markov length work as a diagnostic measure for a ``gap'' of mixed states, 
the characterization of non-trivial mixed state phases and their mixed state phase transitions.

The rest of this paper is organized as follows. 
In Sec.~II, we start to introduce the CMI and the second R\'{e}nyi CMI and review the previous work \cite{Sang2025}, in which the notion of the phase equivalence of mixed state 
by considering the Petz recovery map is shown. 
In Sec.~III, we introduce the doubled Hilbert space formalism. 
Based on this formalism, we show the efficient numerical methods for the second R\'{e}nyi CMI, which can be smoothly carried out by employing an MPS numerical calculation library \cite{TeNPy,Hauschild2024}.
In Sec.~VI, we perform the systematic numerical calculations of the second R\'{e}nyi CMI by using the MPS and the filtering method proposed in 
the author's previous work \cite{Orito2025} to the MPS for various decoherence parameters. 
Here, we find that the second R\'{e}nyi CMI classifies non-trivial mixed states and state phase transitions between them.
Section VIII is devoted to summery and conclusion.

\section{Markov length and its second R\'{e}nyi version}
In this section, we introduce the CMI and Markov length and explain their physical significance as a measure of the system's recoverable properties under decoherence. 
Throughout this study, we focus on the second R\'{e}nyi version of them, which is used in Ref.~\cite{Zhang2024strong}. 
The CMI and its Markov length determine whether there is a recovery map, the existence of which gives the criterion of mixed-state phase equivalence. 
In this section, we review the above content following Ref.~\cite{Sang2025}.

\subsection{CMI and its Markov length}
Quantum informational correlation of the system classifies non-trivial mixed states and gives the notion of the ``gap'' in mixed states analogous to the one of pure states. 
The CMI is one of the important candidates for measuring such a correlation. 

\begin{figure}[t]
\begin{center} 
\includegraphics[width=8cm]{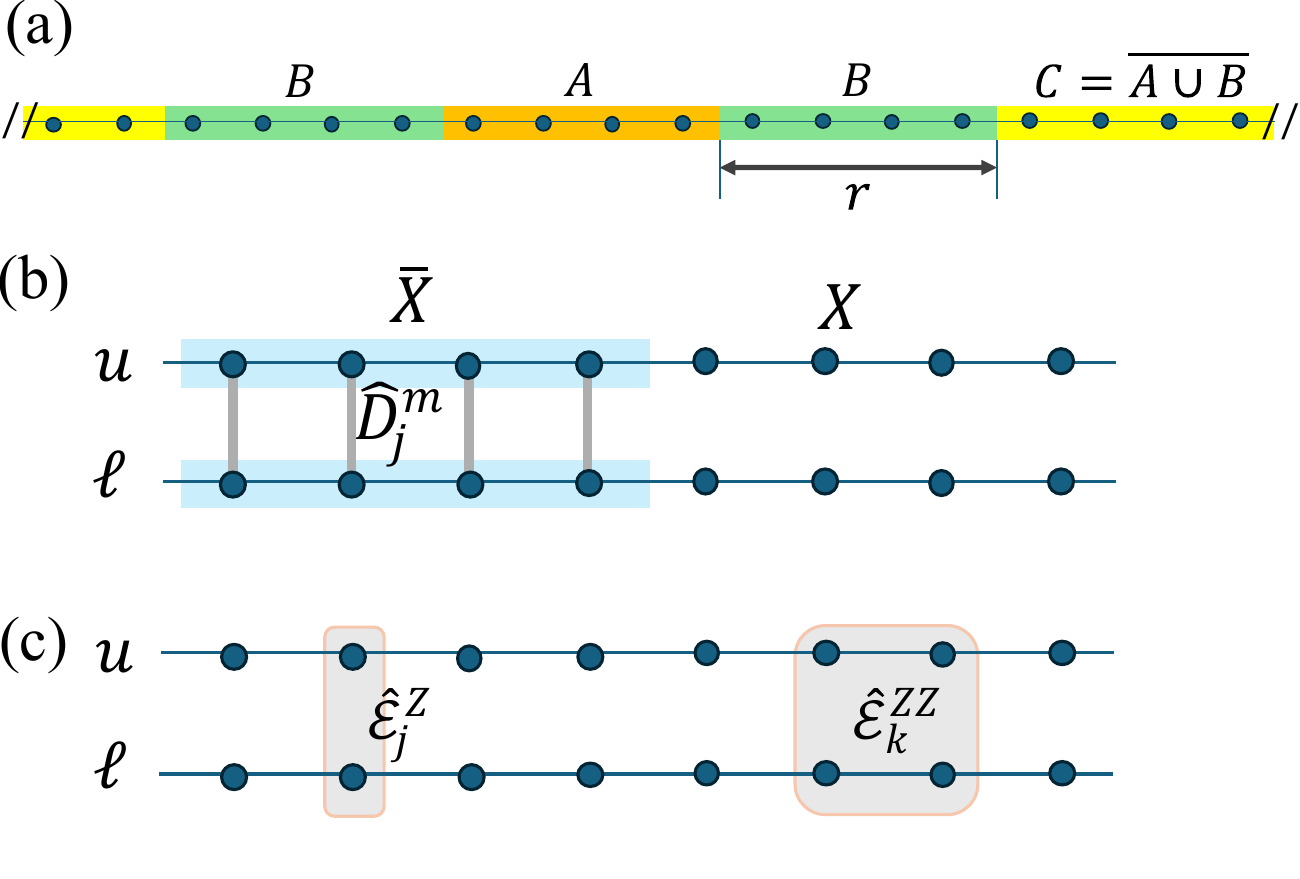}  
\end{center} 
\caption{Schematic image of the system: 
(a) Tripartite for the conditional mutual information. Periodic boundary conditions are imposed.
(b) Image of the system in the doubled Hilbert space that is composed of the upper ($u$) and lower ($\ell$) spin-$1/2$ chains. 
The decoherence operators are regarded as the couplings between the upper and lower chains.
(c) In the doubled Hilbert space picture, the partial trace for the subsystem $\bar{X}$ (${\rm Tr}_{\bar{X}}$) is converted to the maximum depolarizing interchain interaction. 
}
\label{Fig_system}
\end{figure}

The CMI can be evaluated by partitioning the system into three subsystems \cite{Sang2025}. 
Throughout this work, we consider one-dimensional (1D) spin-$1/2$ systems (1D qubits) under periodic boundary conditions, 
partitioned by three subsystems labeled by $A$, $B$ and $C$. 
The subsystem $A$ is sandwiched by the subsystem $B$ and the subsystem $C$ is the complementary set, $C=\overline{A \cup B}$. 
The schematic is shown in Fig.~\ref{Fig_system} \changedd{(a)}. 
We set the total system size $L$ and the lengths of the subsystems corresponding to the number of spins in these subsystems such as $|A|=N_A$, $|B|=2r$ and $|C|=L-N_A-2r$. 
In particular, $r$ is the size of the regime adjacent to $A$, which is one of two portions of the subsystem $B$ as shown in Fig.~\ref{Fig_system} (a). 
Then, by using this partition of the system with $N_A$ and $L$ fixed, the CMI for a pure or mixed state described by $\rho$ is given as,
\begin{eqnarray}
I(A:C|B)(r)&\equiv&I(A:BC)(r)-I(A:B)(r)\nonumber\\
&=&S_{BC}-S_{ABC}-S_B+S_{AB},
\label{CMI0}
\end{eqnarray}
where $S_{X}$ is the von Neumann entanglement entropy of the subsystem $X$, $S_{X}=-{\rm Tr}_{X}[\rho_X\ln \rho_X]$ with $\rho_X={\rm Tr}_{\bar{X}}[\rho]$. 
Intuitively, the CMI is a measure observing how much the correlations between subsystems $A$ and $C$ are screened or separated by the intermediate region $B$.

The behavior of the CMI depends on the magnitude of the regime $B$, i.e., $r$. 
If $r$-dependence of the CMI exhibits an exponential decay, then the Markov length denoted by $\xi_{M}$ is defined as, 
\begin{eqnarray}
I(A:C|B)(r) \sim e^{-r/\xi_{M}}.
\label{CMI_exp_form}
\end{eqnarray}
The inverse of $\xi_M$ can be regarded as a ``gap'' of mixed states that has information-theoretical meaning~\cite{Sang2025}. 
The weak exponential decay of the CMI means that $\xi_M$ is large and the gap of the state is small, and therefore, 
the divergence of $\xi_M$ indicates a phase transition at that point. 

In this study, we consider $n$-th R\'{e}nyi mutual information, which is defined through the R\'{e}nyi entropy as follows,
\begin{eqnarray}
I^{(n)}(A,B)&\equiv& S^{(n)}_A+S^{(n)}_B-S^{(n)}_{AB},\\
S^{(n)}_X&\equiv& \frac{1}{1-n}\log [\Tr_X \rho^{n}_X].
\label{2nd_RE}
\end{eqnarray}
Then, in the tripartite system that we consider, the $n$-th CMI is given by
$$
I^{(n)}(A:C|B)\equiv I^{(n)}(A,BC)-I^{(n)}(A,B). 
$$

In later discussions, we focus on the $n=2$ case, the second R\'{e}nyi CMI \cite{Zhang2024strong}. 
By means of the second R\'{e}nyi CMI, an extended Markov length can be introduced if the CMI shows an exponential decay as a function of $r$,  
\begin{eqnarray}
I^{(2)}(A:C|B)(r) \sim e^{-r/\xi^{(2)}_{M}}.
\label{CMI2_relation}
\end{eqnarray}
We sometimes call $\xi^{(2)}_M$ the second-order Markov length.
From previous numerical works \cite{Li_2019,Zabalo_2020} and an argument in \cite{Micallo2020}, we expect the following relation 
\begin{eqnarray}
I(A:C|B)=\beta I^{(2)}(A:C|B),
\label{R2_CMI_exp}
\end{eqnarray}
where $\beta$ is some positive factor.

\subsection{Recovery map and phase equivalence for mixed states}

In this section, we give a formal and qualitative discussion on the recovery map, which is useful for understanding the mixed-state phase equivalence 
between two mixed states, $\rho_1$ and $\rho_2$.

In~\cite{Sang2025}, a criterion has been proposed: 
Let us assume that $\rho_{2}$ is connected to $\rho_1$ by a certain quantum channel. 
If there exists a recovery quantum channel connecting two mixed states from $\rho_{2}$ to $\rho_1$, these states are regarded as equivalent. 

Existence of a sufficient recovery channel is closely related to the behavior of the CMI of Eq.~(\ref{CMI0}). 
Here, let us assume that state $\rho_2$ is derived by applying a local channel $\mathcal{E}_{loc}$ to state $\rho_1$ as $\rho_2=\mathcal{E}_{loc}[\rho_1]$, 
where we impose the constraint that the channel $\mathcal{E}_{loc}$ locally acts within the regime of the subsystem $A$. 
In general, based on the information of $A\cup B$ subsystem, $\rho_{1,AB}$, the channel $\mathcal{E}_{loc}$ and its adjoint recovery map $\tilde{\mathcal{E}}_{loc}$, namely Petz map, 
can be constructed \cite{Petz1988,Junge2018,Sang2025}. [The detailed formula is shown in \cite{Petz1988,Junge2018}.]
Then the recovery map $\tilde{\mathcal{E}}$ acts to the state $\rho_2$ and $\rho_2$ returns to the state $\rho_1$ if the map $\tilde{\mathcal{E}}$ works efficiently. 
It was proved that the efficiency is bounded by the behavior of the CMI of $\rho_1$ as follows~\cite{Sang2025};  
\begin{eqnarray}
\frac{1}{2 \ln 2} \left\| \tilde{\mathcal{E}}_{loc}[\rho_2] - \rho_1 \right\|_1^2 
&\leq& I_{\rho_1}(A : C | B) - I_{\rho_2}(A:C|B)\nonumber\\
&\leq& I_{\rho_1}(A:C|B)\nonumber\\
&=&\frac{1}{\beta}I^{(2)}_{\rho_1}(A:C|B). 
\label{E_CMI_cond}
\end{eqnarray}
This means that if the CMI of the state $\rho_1$ exhibits exponential decay with respect to $r$ (the form of Eq.~(\ref{CMI_exp_form})), 
then the influence of the channel $\mathcal{E}_{loc}$ acting within the subsystem $A$ remains within the region encompassing subsystems $A$ and $B$. 
Indeed, the error of the recovery map decays exponentially with respect to the width of the subsystem B, $r$.
In this case, the recovery performance of the Petz map constructed from the original information in the subsystems $A$ and $B$ is expected to be highly sufficient.
In the last line of Eq.~(\ref{E_CMI_cond}), we have used Eq.~(\ref{R2_CMI_exp}). 
We expect that the second R\'{e}nyi CMI gives a qualitative bound to the sufficiency of the recovery map.

We further consider the case where multiple local channels act on $\rho_1$ and they are composed of two kinds of local channels, each of which is
similar to $\mathcal{E}_{loc}$ with different application locations. 
We assume that local channels such as $\mathcal{E}_{loc}$ are spatially separated with each other at least a distance $2r'$ and sequentially applied to $\rho_1$ in a localized manner. 
We regard the global channel as a quantum circuit~\cite{Sang2025}. 
The recovery map for such a global circuit channel can be constructed as a composition of the recovery maps associated with each local channel. 
Moreover, the efficiency of the recovery map for this global channel can be also governed by the behavior of the CMI of $\rho_1$. 
Formally, the global channel is written as $\mathcal{E}_g\equiv \prod^{N-1}_{k=0}\mathcal{E}_{loc,k}$. 

Then, let us consider the CMI for $\rho_1$ parameterized as Eq.~(\ref{CMI_exp_form}). 
There exists approximate local recovery maps $\tilde{\mathcal{E}}_{loc,k}$ for each $\mathcal{E}_{loc,k}$. 
Consequently, the global recovery map for $\mathcal{E}_g$ is given by $\tilde{\mathcal{E}}_g=\prod_{k'}\tilde{\mathcal{E}}_{loc,k'}$,
where the application order of local channels is inverse.
Qualitatively, the efficiency of the recovery map $\tilde{\mathcal{E}}_g$ can be estimated by the behavior of the CMI of $\rho_1$ \cite{Sang2025},
\begin{eqnarray}
&&\left\| \tilde{\mathcal{E}}_{g}[\mathcal{E}_{g}[\rho_1]] - \rho_1 \right\|_1^2 \lesssim \sum_{k}
\left\| \tilde{\mathcal{E}}_{loc}[\rho_2] - \rho_1 \right\|_1^2\nonumber\\
&&\leq \sum_{k} \biggr[ (2 \ln 2) I_{\rho_1}(A:C|B)\biggl]^{\frac{1}{2}} 
=(\sqrt{2\ln 2})\alpha(L)e^{-r'/2\xi_{M,\rho_1}}. \nonumber \\
\label{E_CMI_cond_2}
\end{eqnarray}
Here, we assume that (i) each $\mathcal{E}_{loc,k}$ is similar to $\mathcal{E}_{loc}$, 
(ii) the system's Markov length $\xi_M$ does not significantly vary during the action of each local channel and 
keeps property close to the Markov length of $\rho_1$, $\xi_{M,\rho_1}$.
We also expect that the prefactor $\alpha(L)$ of the CMI does not exhibit divergences that would obstruct the exponential decay of the CMI with respect to $r'$. 
Thus, the above relation says that if the CMI has a sufficiently small Markov length, the error of the global recovery map is small. 
Then the two states $\rho_1$ and $\mathcal{E}_g[\rho_1]$ are invertible, implying that they belong to the same mixed state phase.

From the above consideration, if the target state's Markov length is short and remains finite throughout its continuous deformation via some quantum channel---i.e., 
the Markov length does not diverge and the gap does not close---then we expect that a highly efficient recovery channel exists. 
In such cases, the two mixed states connected by this continuous channel can be regarded as equivalent.

If we observe divergence of the Markov length $\xi_{M,\rho_1}$ at a certain point, on the other hand, it implies that the mixed state is unstable, and it
tends to change drastically through a mixed state phase transition, which is analogous to a gap closing on a phase transition in the pure state framework.

So far we have considered the CMI defined by the von Neumann entropy. 
In what follows, we focus on the second R\'{e}nyi CMI, as it is technically more tractable by the MPS numerical simulation as shown in the later section. 
We expect that the second R\'{e}nyi CMI can give the essential insight into the recovery bound relation of Eq.~(\ref{E_CMI_cond_2}). 
The second R\'{e}nyi CMI and its Markov length play a key role in classifying mixed states with respect to phase equivalence and characterizing mixed state phase transitions 
since the CMI and the second one are expected to be related in Eq.~(\ref{R2_CMI_exp}).

Similarly to the original CMI \cite{Sang2025,Negari2024}, we expect that the second R\'{e}nyi CMI behaves as 
\begin{eqnarray}
I^{(2)}(A : C \mid B) \simeq 
\begin{cases}
e^{-r / \xi^{(2)}_M(p)} & p < p_c \\
r^{-\alpha^{(2)}} & p = p_c \\
e^{-r / \xi^{(2)}_M(p)} & p > p_c,
\end{cases}
\end{eqnarray}
where $p$ is a system parameter, and $\alpha$ is a critical exponent if some mixed state phase transition takes place at a critical parameter point $p_c$. 
In the rest of the present work, we numerically investigate whether the above behavior, especially the exponential behavior, 
emerges and if so, we estimate the second R\'{e}nyi Markov length for two concrete systems under decoherence.

\section{Efficient calculation methods of the second R\'{e}nyi CMI}

In this section, we explain efficient numerical methods to calculate the second R\'{e}nyi CMI for mixed states by using the doubled Hilbert space formalism. 

In general, a mixed state density matrix (the decohered state) $\rho\in \mathcal{H}$, where $\mathcal{H}$ is a target Hilbert space, 
can be efficiently treated as a pure state by the doubled Hilbert space formalism. 
In this formalism, the target Hilbert space $\mathcal{H}$ is doubled as $\mathcal{H}_{u}\otimes \mathcal{H}_{\ell}$, 
where the subscripts $u$ and $\ell$ refer to the upper and lower Hilbert spaces corresponding to the ket and bra states of mixed state density matrix, respectively. 
Then, the 1D system we consider can be regarded as a ladder system shown in Fig.~\ref{Fig_system} (b).
Density matrix $\rho$ is vectorized as $\rho \longrightarrow |\rho\rangle\rangle\equiv \frac{1}{\sqrt{\dim[\rho]}}\sum_{k}|k\rangle\otimes \rho|k\rangle$, 
where $\{|k\rangle \}$ is an orthonormal set of bases in the Hilbert space $\mathcal{H}$. 
This is the Choi-Jamio\l kowski isomorphism \cite{Choi1975,JAMIOLKOWSKI1972}, and 
the state $|\rho\rangle\rangle$ is in the doubled Hilbert space $\mathcal{H}_u\otimes \mathcal{H}_{\ell}$.

We show how to calculate the second R\'{e}nyi entropy corresponding to $n=2$ case of Eq.~(\ref{2nd_RE}) for the vector $|\rho\rangle\rangle$. 
We first divide each Hilbert space into $\mathcal{H}_{u(\ell)}=\mathcal{H}_{X,u(\ell)}\otimes \mathcal{H}_{\bar{X},u(\ell)}$, 
where one subsystem is represented $X$ and the other is $\bar{X}$. 
We calculate the second R\'{e}nyi entropy for the subsystem $X$, $S^{(2)}_X$.
To this end, we trace out the degrees of the freedom in the subsystem $\bar{X}$ and obtain the trace of the squared reduced density matrix 
$\rho^2_{X}\equiv ({\rm Tr}_{\bar{X}}\rho)^{2}$. 
For the supervector $|\rho\rangle\rangle$, this manipulation can be carried out by applying a suitable maximal depolarization channel to the subsystem $\bar{X}$ \cite{zhang2025_SP}.

The maximal depolarization channel \cite{Nielsen2011} is represented by an operator (A brief explanation of which is given in Appendix A) 
acting on the vector $|\rho\rangle\rangle$ \cite{Lee2025}, which is given by
\begin{eqnarray}
\hat{D}_{\bar{X}}&=&\prod_{j\in \bar{X}}\hat{D}^{m}_j,\\
\hat{D}^{m}_j&=&\frac{1}{4}\biggr[\hat{I}_{j,u} \otimes \hat{I}_{j,\ell}+\hat{X}_{j,u} \otimes \hat{X}_{j,\ell}-\hat{Y}_{j,u}\otimes \hat{Y}_{j,\ell}\nonumber\\
&&+\hat{Z}_{j,u} \otimes \hat{Z}_{j,\ell}\biggr],
\end{eqnarray}
where $\hat{I}_{j,u(\ell)}$ is an identity operator for site-$j$ vector space in $\mathcal{H}_{u(\ell)}$, $Z(X,Y)_{j,u(\ell)}$ is Pauli-$Z$($X$,$Y$) operator at site $j$. 
The channel operator $\hat{D}_{\bar{X}}$ is regarded as a coupling between the system $u$ and $\ell$, as shown in Fig.~\ref{Fig_system} (b). 
Then, the application of $\hat{D}_{\bar{X}}$ leads to the following identity~\cite{zhang2025_SP}, 
\begin{eqnarray}
\hat{D}_{\bar{X}}|\rho\rangle\rangle=|\frac{{\bf I}_{\bar{X}}}{d_{\bar{X}}}\otimes \rho_{X}\rangle\rangle.
\label{MD_1}
\end{eqnarray}
where $d_{\bar{X}}$ is the number of the degree of freedom of the subsystem $\bar{X}$ and 
$|\frac{{\bf I}_{\bar{X}}}{d_{\bar{X}}}\otimes \rho_{X}\rangle\rangle \longleftrightarrow \frac{{\bf I}_{\bar{X}}}{d_{\bar{X}}}\otimes [{\rm Tr}_{\bar{X}}\rho]$, 
that is, the maximal depolarizing channel projects the state of the subsystem $\bar{X}$ onto a state at infinite temperature, as a result,
all information in the subsystem $\bar{X}$ is swept out.

Furthermore since inner product of the doubled Hilbert space vectors $\langle\langle A|B\rangle\rangle$ is nothing but $\Tr[A^\dagger B]$ \cite{Ma2024_double}, 
the norm of the state $\hat{D}_{\bar{X}}|\rho\rangle\rangle$ is related to $S^{(2)}_X$ as follows,
\begin{eqnarray}
\langle\langle \frac{{\bf I}_{\bar{X}}}{d_{\bar{X}}}\otimes \rho_{X} |\frac{{\bf I}_{\bar{X}}}{d_{\bar{X}}}\otimes \rho_{X}\rangle\rangle 
=\frac{1}{d_{\bar{X}}}{\rm Tr}_{X}[\rho^2_{X}].
\label{norm_cal}
\end{eqnarray}
From the above relation, $S^{(2)}_X$ can be obtained by the norm of the state $\hat{D}_{\bar{X}}|\rho\rangle\rangle$,
\begin{eqnarray}
S^{(2)}_X=-\log\biggr[d_{\bar{X}}\langle\langle \frac{{\bf I}_{\bar{X}}}{d_{\bar{X}}}\otimes \rho_{X} |\frac{{\bf I}_{\bar{X}}}{d_{\bar{X}}}\otimes \rho_{X}\rangle\rangle \biggl].
\end{eqnarray}

The operation of Eq.~(\ref{MD_1}) and the norm calculation in Eq.~(\ref{norm_cal}) can be easily performed by the MPS numerical-based filtering methods
presented in previous studies \cite{Orito2025,kuno_2025,kuno_2025_v1}.
Thus, by setting the subsystem $X$ suitably, we can efficiently calculate the second R\'{e}nyi CMI, $I^{(2)}(A:C|B)$.

\begin{figure}[b]
\begin{center} 
\includegraphics[width=7.5cm]{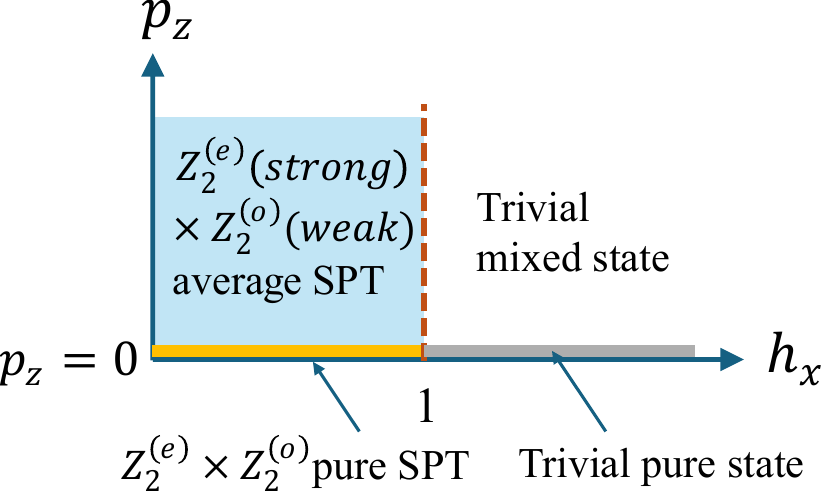}  
\end{center} 
\caption{Schematic image of the phase diagram on $p_z$-$h_x$ parameter space.
}
\label{PD_cluster}
\end{figure}
\begin{figure}[t]
\begin{center} 
\includegraphics[width=8.7cm]{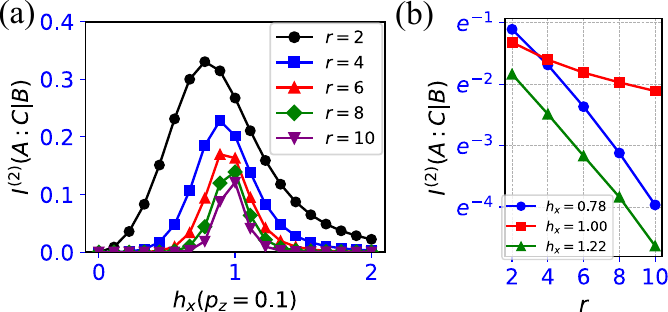}  
\end{center} 
\caption{(a) $h_x$-dependence of $I^{(2)}(A:C|B)$ with $p_z=0.1$ for various $r$'s. 
(b) $r$-dependence of $I^{(2)}(A:C|B)$ for three typical $h_x$'s in the vicinity of the phase transition. 
Numerically estimated $\xi^{(2)}_M$ are $\sim 2.801$, $\sim 10.204$, and $\sim 2.881$ at $h_x= 0.78$, $1.0$ and $1.22$.
}
\label{Fig_CMI_ZXZ}
\end{figure}
\section{Numerical results}

In what follows, we show numerical calculations of the second R\'{e}nyi CMI, using DMRG and MPS simulations \cite{TeNPy,Hauschild2024} based on the doubled Hilbert space formalism, 
where we use the TeNPy library \cite{TeNPy,Hauschild2024}. 
We consider two kinds of 1D systems under decoherence and observe the decohered states emerging from pure states and calculate $r$ dependence of the CMI 
to observe the behavior of the second R\'{e}nyi Markov length. 
In all numerical simulations, we set $|A|=N_A=4$, $|B|=2r$ and $L$ satisfying $|C|=L-N_A-2r=r$.

\subsection{Cluster model under odd-site $Z$ decoherence}

The first target model is the cluster model with on-site magnetic field \cite{Raussendorf2001,Else2012}, the Hamiltonian of which is given by
\begin{eqnarray}
H_{\rm cl}=\sum^{L-1}_{j=0}[-Z_{j-1}X_{j}Z_{j+1}+h_xX_j],
\label{Hcl}
\end{eqnarray}
where $h_x$ is a magnetic field. 
For $|h_x|<1$, the ground state is a $Z^{(e)}_2\times Z^{(o)}_2$ SPT state, the symmetry generators of which are
$\prod_{j\in even}X_j$ and $\prod_{j\in odd}X_j$.

We study how the ground state of $H_{\rm cl}$ evolves under odd-site $Z$-decoherence, the Choi operator representation of which is given as 
\begin{eqnarray}
\hat{\mathcal{E}}_{Z}&=&\prod_{j\in odd}\hat{\mathcal{E}}_{Z,j},\nonumber\\
\hat{\mathcal{E}}_{Z,j}&=&\biggr[(1-p_z)\hat{I}_{j,u} \otimes \hat{I}_{j,\ell}+p_z Z_{j,u}\otimes Z_{j,\ell}\biggl],
\end{eqnarray}
where $p_z$ is strength of the decoherence, $0\leq p_z\leq 1/2$. 
The $p_z=1/2$ case corresponds to non-selective projective measurement, and  
for $p_z \neq 0$, the channel $\mathcal{E}_{Z}$ changes the ground states to a mixed state. 
Very recently, it was found that the decohered mixed state emerging from the SPT ground state retains an SPT order and is called an ASPT phase protected 
by the weak $Z^{(o)}_2$ and strong $Z^{(e)}_2$ symmetry \cite{Lee2025,ma2023,ma2024,Ma2024_double}. 
For $p_z=0$ case, it is known that a phase transition takes place from the SPT state to the trivial pure state by increasing $h_x$.
We expect that this transition remains even for finite $p_z$, but the symmetry for the SPT order reduces to the weak $Z^{(o)}_2$ and strong $Z^{(e)}_2$ symmetry \cite{ma2024,Ma2024_double,Lessa_PRXQ}. 
[The definition of the strong and weak symmetry is given in Refs.~\cite{Buca_2012,groot2022}.]
The schematic image of the global phase diagram is shown in Fig.~\ref{PD_cluster}. 

We observe the behavior of the second R\'{e}nyi CMI for the decohered state for a fixed $p_z$ with varying the initial pure state by changing $h_x$. 
That is, we observe the second R\'{e}nyi CMI of the decohered state $\hat{\mathcal{E}}_{Z}|\rho_{\rm cl}(h)\rangle\rangle$. 
Here, $|\rho_{\rm cl}(h)\rangle\rangle$ is the supervector representation of the initial density matrix (pure state) $\rho_{\rm cl}(h)=|\psi_{\rm cl}(h)\rangle\langle \psi_{\rm cl}(h)|$ 
where $|\psi_{\rm cl}(h)\rangle$ is the unique ground state of $H_{\rm cl}$. 
The state is prepared by DMRG calculation \cite{DMRG_condition}.

Figure \ref{Fig_CMI_ZXZ}(a) displays $h_{x}$-dependence of the second R\'{e}nyi CMI with $p_{z}=0.1$ for various $r$'s. 
The peaks appear at a specific value of $h_x$ for each $r$. 
This indicates the existence of a mixed state phase transition from the ASPT to the trivial mixed phase with increasing $h_x$. 
To elucidate this result, we further observe $r$-dependence of the second R\'{e}nyi CMI for $h_{x}=0.78$, $1.0$ and $1.22$ around peaks. 
The data is displayed in Fig.~\ref{Fig_CMI_ZXZ}(b). 
The two cases $h_{x}=0.78$ and $1.22$ exhibit a clear exponential decay as expected in Eq.~(\ref{CMI2_relation}). 
Therefore, the second R\'{e}nyi Markov length $\xi^{(2)}_M$ is small for the both cases.
On the other hand, the case $h_{x}=1$ exhibits very weak exponential decay. 
By using an exponential fitting with the fitting function $I^{(2)}(A:C|B)(r)=e^{-c_0r}+c_1$, 
numerically extracted values of $\xi^{(2)}_M(=c^{-1}_0)$ are $\sim 2.801$, $\sim 10.204$, and $\sim 2.881$ at $h_x= 0.78$, $1.0$, and $1.22$, respectively. 
Thus, we expect that at an expected phase transition point $h_x\sim 1.0$, the second R\'{e}nyi Markov length $\xi^{(2)}_M$ diverges in the thermodynamics limit. 
On the other hand, from the result of the second R\'{e}nyi Markov length $\xi^{(2)}_M$, the average SPT and trivial mixed states are expected to have a large mixed state ``gap''. 

These results indicate that the average SPT regime ($h_x < 1$) has a small second Markov length. 
Thus, as discussed in Sec.~II B, if we pick up two mixed states in this regime and if there exists a suitable local channel such as $\mathcal{E}_g$ in Sec.~II B, 
which connects the two states, there can be an efficient recovery map $\tilde{\mathcal{E}}_g$. 
This discussion is also applicable to the regime of the trivial mixed state ($h_x > 1$).

\subsection{Transverse field Ising chain under $ZZ$-$X$ decoherence and SWSSB phase transition}
We turn to another numerical demonstration. The second target system is 1D TFIM, the Hamiltonian of which is given by
\begin{eqnarray}
H_{\rm TFIM}=-\sum^{L-1}_{j=0}[J_{zz}Z_jZ_{j+1}+h_x X_j],
\end{eqnarray}
where $J_{zz}$, $h_x>0$ are parameters. 

\begin{figure}[b]
\begin{center} 
\includegraphics[width=6cm]{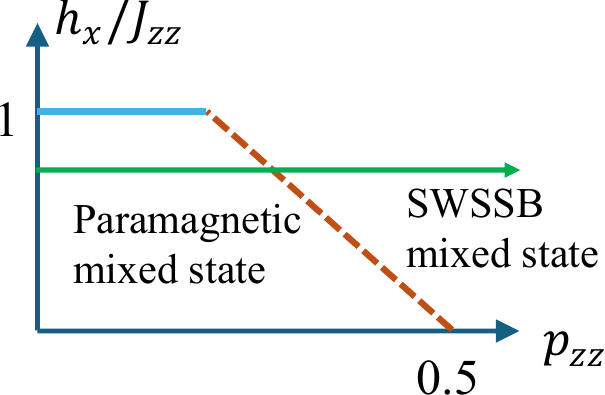}  
\end{center} 
\caption{Schematic image of the phase diagram on $p_{zz}$-$h_x/J_{zz}$ parameter space.
A similar phase diagram has been investigated in \cite{Orito2025}. The green line is a target parameter sweep in this work.}
\label{PD_qAT}
\end{figure}
\begin{figure}[t]
\begin{center} 
\includegraphics[width=9cm]{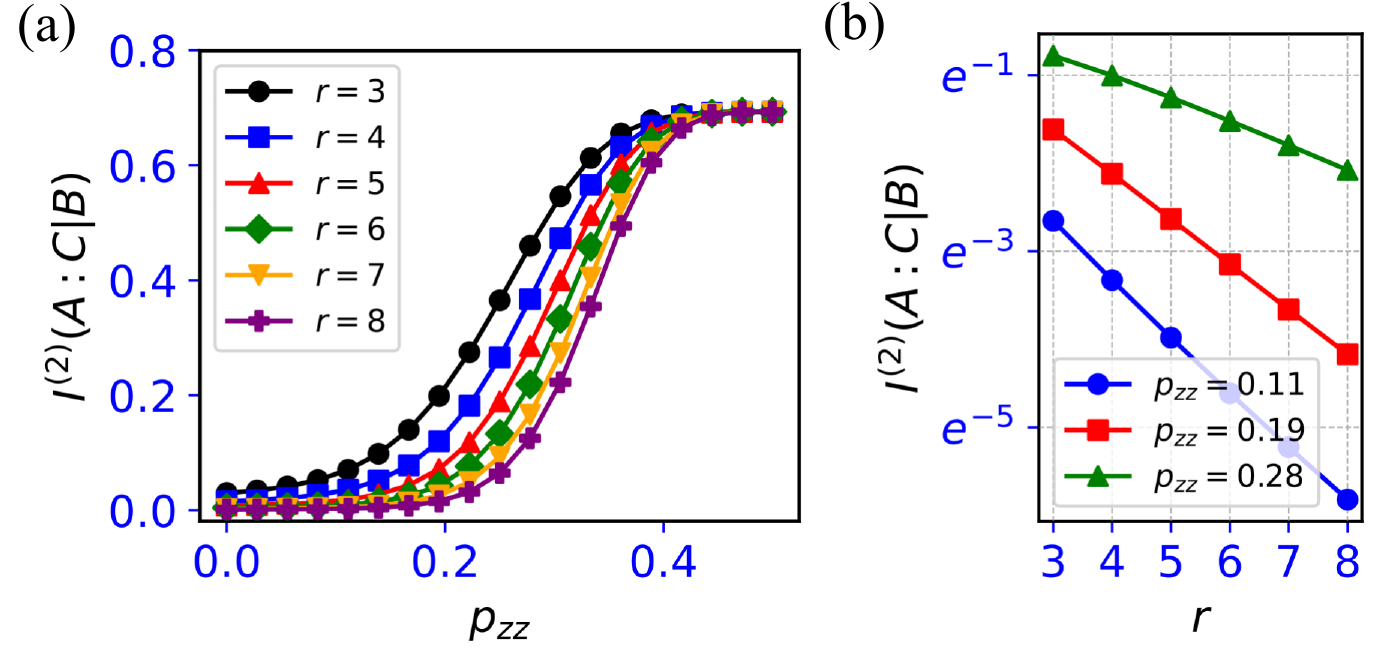}  
\end{center} 
\caption{(a) $p_{zz}$-dependence of $I^{(2)}(A:C|B)$ for various $r$'s. 
(b) $r$-dependence of $I^{(2)}(A:C|B)$ for three typical $p_{zz}$'s. 
Numerically estimated $\xi^{(2)}_M$ are $\sim 1.580$, $\sim 1.954$, and $\sim 3.825$ at $p_{zz}=0.11$, $0.19$, and $0.28$.
}
\label{Fig_SWSSB}
\end{figure}

For the ground state of $H_{\rm TFIM}$, we apply the following two types of decoherence, the Choi representations of which are given as,
\begin{eqnarray}
\hat{\mathcal{E}}_{ZZ}&=&\prod^{L-1}_{j=0}\biggr[(1-p_{zz})\hat{I}_{j,u}^* \otimes \hat{I}_{j,\ell}\nonumber\\ &&+p_{zz}Z_{j,u}^*Z_{j+1,u}^*\otimes Z_{j,\ell}Z_{j+1,\ell}\biggl],\\ 
\hat{\mathcal{E}}_{X}&=&\prod^{L-1}_{j=0}\biggr[(1-p_{x})\hat{I}_{j,u}^* \otimes \hat{I}_{j,\ell} +p_{x}X_{j,u}^*\otimes X_{j,\ell}\biggl],
\end{eqnarray}
where, $p_{zz(x)}$ is a strength of the decoherence, $0\leq p_{zz(x)}\leq 1/2$. 
The decoherence can be regarded as the inter-chain coupling in the doubled-Hilbert state as shown in Fig.~\ref{Fig_system} (c). 

In what follows, we fix $h_x=1, J_{zz}=0.8$ and consider the decohered state 
$\hat{\mathcal{E}}_{X}\hat{\mathcal{E}}_{ZZ}|\rho_g\rangle\rangle$ \footnote{The application order in $\hat{\mathcal{E}}_{X}$ and $\hat{\mathcal{E}}_{ZZ}$ is commutative.}. 
Here, $|\rho_g\rangle\rangle$ is the supervector representation of the initial density matrix (pure state) $\rho_{g}=|\psi_{g}\rangle\langle \psi_{g}|$ where $|\psi_{g}\rangle$ is 
the unique ground state of $H_{\rm TFIM}$. 
In the practical calculation, the state is prepared by DMRG calculation.
We further consider a single-parameter decoherence, where we sweep $p_{zz}$ with controlling $p_x=1/2-(1/2)(1-2p_{zz})^{h_{x}/J_{zz}}$. 
In the previous work \cite{Orito2025}, under this parameter sweep, we found two mixed phases, paramagnetic mixed and SWSSB phases, 
the schematic image of the phase diagram is shown in Fig.~\ref{PD_qAT}.
A mixed state phase transition takes place between these two mixed phases. 
In that phase diagram, SWSSB can be characterized by various physical quantities, such as fidelity and R\'{e}nyi-2 correlator, as recently considered in \cite{weinstein2024,Lessa_PRXQ,KOI2024}. We shall re-examine these findings by measuring the second R\'{e}nyi CMI and its Markov length.

Figure \ref{Fig_SWSSB}(a) shows the $p_{zz}$-dependence of the second R\'{e}nyi CMI for various $r$'s. 
The value increases as $p_{zz}$ increases, and the $r$ dependence is getting larger in the region $0.1 \lessapprox p_{zz} \lessapprox 0.2$. 
This implies that in the vicinity of $p_z=0.2$, the drastic change of the second R\'{e}nyi Markov length occurs.
Then, the CMI saturates at $p_z \sim 0.4$.
This behavior indicates the existence of the mixed state phase transition between the paramagnetic mixed and SWSSB states. 

We further find that the value of the second R\'{e}nyi CMI approaches $\ln 2$ for $p_{zz}\to 1/2$. As for Ref.~\cite{Lessa_PRXQ} it has been predicted that the CMI of a SWSSB state takes the universal value, $\ln 2$. 
Our numerical calculation implies that the second R\'enyi CMI similarly takes the value $\ln 2$ in the deep SWSSB phase. 
Furthermore, as shown in Appendix~C, this value coincides with that of a typical SWSSB state firstly suggested as the simplest SWSSB phase \cite{Lessa_PRXQ,sala2024} 
in the stabilizer state limit, indicating that it is likely a universal feature.

To elucidate this result shown in Fig.~\ref{Fig_SWSSB} (a), we further observe $r$-dependence of the second R\'{e}nyi CMI for $p_{zz}=0.11$, $0.19$ and $0.28$ around a transition point. 
The data is displayed in Fig.~\ref{Fig_SWSSB}(b). 
Clear exponential decays as expected in Eq.~(\ref{CMI2_relation}) can be verified but some background constant appears and increases as $p_{zz}$ increases. 
The decay as a function of $r$ is weaker for larger $p_{zz}$, meaning that 
the second R\'{e}nyi Markov length $\xi^{(2)}_M$ gets larger for larger $p_{zz}$. 

For these data, we apply the same exponential fitting to the former case. 
We numerically estimated the second R\'{e}nyi Markov length: $\xi^{(2)}_M(=c^{-1}_0)$ are $\sim 1.580$, $\sim 1.954$, and $\sim 3.825$ at $p_{zz}=0.11$, $0.19$, and $0.28$, respectively. 
This fitting result is a signal that the mixed state changes into the SWSSB state as increasing $p_{zz}$. 
From this fitting observation and the results of Fig.~\ref{Fig_SWSSB}, the $r$-dependence of the second R\'{e}nyi CMI vanishes in the limit $p_{zz}\to 1/2$, i.e.,
its exponential decaying behavior as a function of $r$ disappears. 
This means that the Markov length is infinite in the whole region of the SWSSB phase and the second R\'{e}nyi CMI takes a constant value. 
This observation is consistent with the recent argument of Markov length in the SWSSB phase in \cite{Lessa_PRXQ}.

To summarize our claim herein, the transition from the paramagnetic state to the SWSSB state can be characterized from the viewpoint of the Markov length: 
As the system evolves from the paramagnetic state toward the SWSSB transition point, the Markov length increases. 
Within the SWSSB phase, the exponential decaying behavior of the second R\'{e}nyi CMI is no longer observed and the second R\'{e}nyi CMI takes a $r$-independent constant inducing the breakdown of the exponential $r$-dependent form of it, and the second R\'{e}nyi Markov length becomes infinite.
This breakdown of the Markov length picture captures the essential behavior of the SWSSB transition.

We also give some observations for the aspect of the mixed state phase equivalence as discussed in Sec.~II B. 
In this system, the decoherence channel $\hat{\mathcal{E}}_{X}\hat{\mathcal{E}}_{ZZ}$ 
can be regarded as a complex channel $\mathcal{E}_{g}$ considered in Sec.~II B \cite{channel_note}.
In the paramagnetic mixed state regime, the numerical calculation of the second R\'{e}nyi CMI indicates the finite small Markov length. 
Thus, the mixed state $\hat{\mathcal{E}}_{X}\hat{\mathcal{E}}_{ZZ}|\rho_g\rangle\rangle$ with a finite $p_{zz}$ 
not having SWSSB order is equivalent to the pure paramagnetic ground state of $H_{\rm TFIM}$, $|\rho_g\rangle\rangle$ 
since we expect that the small second R\'{e}nyi Markov length preserves the recovery map $\widetilde{\mathcal{E}}_g(=\widetilde{\mathcal{E}_{X}\circ\mathcal{E}_{ZZ}})$ 
for the channel $\hat{\mathcal{E}}_{X}\hat{\mathcal{E}}_{ZZ}$.
However, for the SWSSB regime (large $p_{zz}$ regime), applying the notion of the mixed state phase equivalence to the SWSSB regime should be avoided 
since the second R\'{e}nyi Markov length cannot be well-defined in the regime.

\section{Conclusion}

In this work, we studied the second R\'{e}nyi CMI and the Markov length derived from that measure. 
As demonstrated in Sec.~IV, the second R\'{e}nyi CMI and its Markov length are good measures to classify the regime of a non-trivial mixed state and to locate phase transitions between them, 
similarly to the original CMI and the derived Markov length \cite{Sang2025}. 

We also proposed an efficient calculation method of the second R\'{e}nyi CMI based on the doubled Hilbert space formalism. 
This method is efficiently manipulated in MPS numerical simulation. 
We numerically verified that the second R\'{e}nyi CMI serves as a diagnostic quantity for the gap structure of mixed states.
It also clarifies mixed-state phase transitions through the significant change of the second R\'{e}nyi Markov length as shown by investigating two concrete 1D systems under decoherence. 
For the first model, we showed that the average SPT and trivial mixed phases are precisely characterized by the CMI,
and the phase transition between them is clearly detected by the second R\'{e}nyi Markov length. 
For the second model, we also numerically verified the emergence of the SWSSB and paramagnetic mixed phases and the phase transition between them induced by varying the strength of decoherence. 
Furthermore, we found that the second R\'{e}nyi CMI has a finite value related to the signature of the long-range entanglement in the deep SWSSB phase.
This observation strongly supports the observation and prediction in \cite{Lessa_PRXQ}.

Finally, we shall comment on some future directions. 
An application of the calculation method of the second R\'{e}nyi CMI and its Markov length to some dynamical quantum circuit systems is a promising future direction of study 
such as in \cite{Negari2024}. 
For example, extending the numerical method proposed in Sec.~III to investigation of the many-body negativity is an important future task. 
The negativity and its combination such as topological entanglement negativity \cite{Fan_PRXQ} can be a good measure for detecting and 
classifying non-trivial mixed state such as mixed state topological order \cite{Wang_PRXQ,kuno_2025_v1} and critical transition properties \cite{Zou2023}.

It is also an important issue to clarify the relation between the Markov length and the ordinary correlation length obtained by
correlation functions of physical operators such as order parameters.
To elucidate how the information-theoretic quantities and corresponding physical ones are related with each other is a significant future problem.

\section*{Acknowledgements}
This work is supported by JSPS KAKENHI: JP23K13026(Y.K.) and JP23KJ0360(T.O.). 

\section*{Data availability}
The data that support the findings of this study are available from the authors upon reasonable request.

\renewcommand{\thesection}{A\arabic{section}} 
\renewcommand{\theequation}{A\arabic{equation}}
\renewcommand{\thefigure}{A\arabic{figure}}
\setcounter{equation}{0}
\setcounter{figure}{0}

\appendix
\section*{Appendix}
\subsection*{A.Choi representation for decoherence channel}
Quantum decoherence is expressed in terms of Kraus operators~\cite{lidar2020}; $\mathcal{E}[\rho]=\sum^{M-1}_{\alpha=0}K_{\alpha}\rho K^\dagger_{\alpha}$, 
where $K_\alpha$'s are Kraus operators satisfying $\sum^{M-1}_{\alpha=0}K_{\alpha}K^{\dagger}_{\alpha}=I$.
In the Choi isomorphism, the channel operator is transformed as ~\cite{Nielsen2011,Lee2025} $\mathcal{E}\longrightarrow \hat{\mathcal{E}}=\sum^{M-1}_{\alpha=0}K^*_{\alpha,u}\otimes K_{\alpha,\ell}$.
That is, the quantum channel represented by the Kraus operators becomes generally a non-unitary operator for the mixed state supervector in the doubled Hilbert space. 
Since the channel operator $\hat{\mathcal{E}}$ is not a unitary map in general cases although the channel is a completely positive trace-preserving map, 
the application of the channel operator generally changes the norm of the mixed state supervector. 
In fact, the norm is nothing but the purity of the original mixed state \cite{Ashida2024}.

\subsection*{B. Additional data in Sec.~IV A}
\begin{figure}[t]
\begin{center} 
\includegraphics[width=8cm]{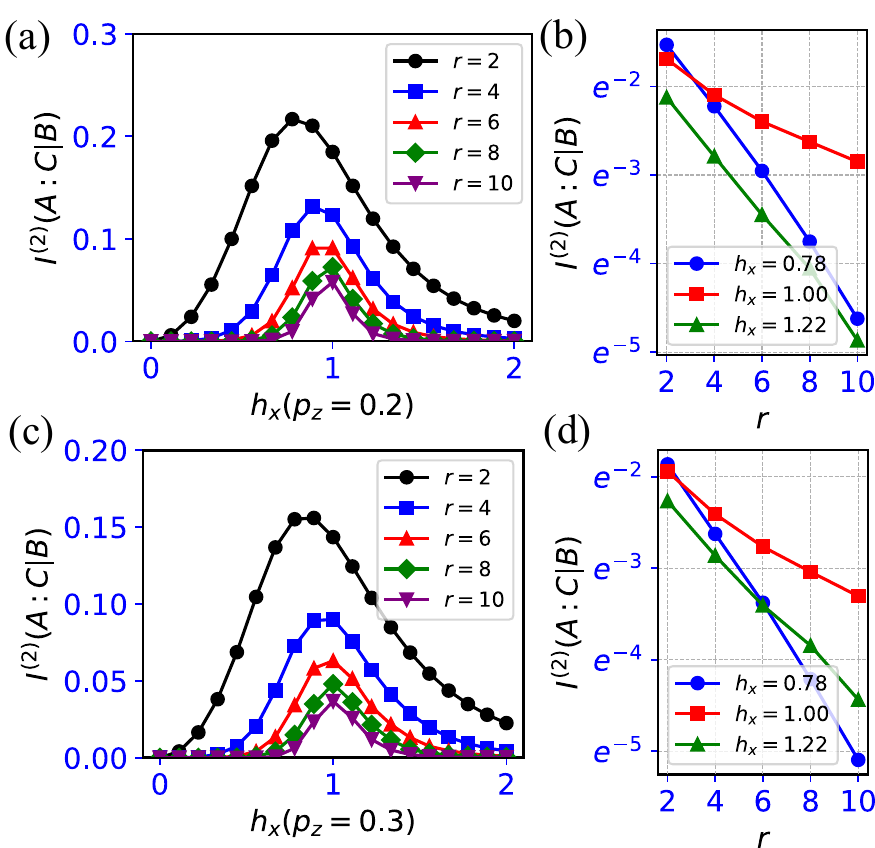}  
\end{center} 
\caption{(a) $h_x$-dependence of $I^{(2)}(A:C|B)$ with $p_z=0.2$ for various $r$'s. 
(b) $r$-dependence of $I^{(2)}(A:C|B)$ for three typical $h_x$'s on and around the transition. 
Numerically extracted $\xi^{(2)}_M$ are $\sim 2.592$, $\sim 7.011$ and $\sim 2.961$ at $h_x= 0.78$, $1.0$ and $1.22$.
(c) $h_x$-dependence of $I^{(2)}(A:C|B)$ with $p_z=0.3$ for different $r$. (d) $r$-dependence of $I^{(2)}(A:C|B)$ for three typical $h_x$'s on and around the transition. Numerically extracted $\xi^{(2)}_M$ are $\sim 2.485$, $\sim 5.960$ and $\sim 3.755$ at $h_x= 0.78$, $1.0$ and $1.22$.
}
\label{Fig_ZXZ_sup}
\end{figure}

We show additional data for the calculation of the second R\'{e}nyi CMI with $p_{z}=0.2$ and $0.3$ in the same setting in Sec.~IV A. 
Figures \ref{Fig_ZXZ_sup}(a)-\ref{Fig_ZXZ_sup}(d) are the numerical results of the second R\'{e}nyi CMI. 
In all displayed data, the same claims as those in Sec.~IV B in the main text are satisfied. 
In addition, we also estimated the second R\'{e}nyi Markov length with $p_{z}=0.2$ and $0.3$. In $p_{zz}=0.2$ case, 
the extracted the second R\'{e}nyi Markov length  $\xi^{(2)}_M(=c^{-1}_0)$ are $\sim 2.592$, $\sim 7.011$, and $\sim 2.961$ at $h_x= 0.78$, $1.0$, and $1.22$. In $p_{zz}=0.3$ case, 
the extracted the second R\'{e}nyi Markov length  $\xi^{(2)}_M(=c^{-1}_0)$ are $\sim 2.485$, $\sim 5.960$, and $\sim 3.755$ at $p_{zz}=0.11$, $0.19$ and $0.28$. 
Both of two cases exhibit large second R\'{e}nyi Markov length $\xi^{(2)}_M$ on $h_x=1$.

\subsection*{C. Second R\'{e}nyi CMI for $J_{zz}\to 0$ and $p_{zz}\to \infty$}
For the TFIM under $ZZ$ and $X$ decoherence shown in Sec.~IV B in the main text, we calculate the CMI for the case $J_{zz}\to 0$ and $p_{zz}\to \infty$. 
We can treat the state and the decoherence in the stabilizer formalism \cite{Nielsen2011,Weinstein_2022}. 
Then, the decohered state can be represented by a set of the generator of the stabilizer. The set of the decohered state is very simple, 
which is only a single generator, $\mathcal{S}=\{\prod^{L-1}_{j=0}X_j\}$ \cite{KOI2024,Lessa_PRXQ,sala2024}, corresponding to a prototypical SWSSB state for $Z_2$ symmetry.

In the stabilizer formalism, each second R\'{e}nyi entropy in the CMI can be obtained by \changedd{\cite{shi2021}}
\begin{eqnarray}
S^{(2)}_X= - (|X|-s_X)\ln 2,
\end{eqnarray}
where $X$ is a subsystem and $|X|$ is the number of spins in the subsystem $X$ and $s_X$ is the number of independent generators of the set of stabilizer group, 
the spatial support of which is within the subsystem $X$. 
Then, the second R\'{e}nyi CMI for the decohered state represented by the stabilizer set $\mathcal{S}$ is $I^{(2)}(A:C|B)= \ln 2$. 

This typical SWSSB state is expected to belong to the same phase as the SWSSB state discussed in the main text. 
Accordingly, the value of the second R\'{e}nyi CMI obtained here coincides with that of the SWSSB phase considered in the main text. 
Thus, the CMI in deep SWSSB states is expected to take a constant value $\ln 2$.

\bibliography{ref}

\begin{thebibliography}{56}
\bibitem [{DMR()}]{DMRG_condition}%
  \BibitemOpen
  \href@noop {} {}\bibinfo {note} {In the DMRG simulations for prepareing initial ground states, we set maximum bond dimension $D=100$-$200$ and truncate the singular value less than $\mathcal{O}(10^{-6})$, and the energy convergence of the iterative DMRG sweep is $\Delta E < \mathcal{O}(10^{-4})$ to obtain the initial MPS ground state.}\BibitemShut {Stop}%
\bibitem [{cha()}]{channel_note}%
  \BibitemOpen
  \href@noop {} {}\bibinfo {note} {The detail form can be constracted by determining the depth of the circuit depending on the strength of the decoherence $p_{zz}$ and the local miltiplied channels in $\hat{\mathcal{E}}_{ZZ(X)}$ can be regarded as a local evolving Lindbladian as shown in \cite{Sang2025}.}\BibitemShut {Stop}%
\end{thebibliography}

\end{document}